\documentstyle[twocolumn,aps,prl]{revtex}

\begin{document}
\draft


\twocolumn[\hsize\textwidth\columnwidth\hsize\csname@twocolumnfalse\endcsname%

\title{NMR study of the $\rm\bf S=1/2$ Heisenberg Ladder
$\rm\bf Cu_2(C_{5}H_{12}N_{2})_2Cl_4$: \\
Quantum Phase Transition and Critical Dynamics}
\author {G. Chaboussant$^{1}$, Y. Fagot-Revurat$^{1}$, M.-H. Julien$^{1}$,
M.E. Hanson$^{1}$ \\
C. Berthier$^{1,3}$, M. Horvati\'{c}$^{1}$, L. P. L\'evy$^{1,2}$ and O. Piovesana$^{4}$}
\address {$^{1)}$ Grenoble High Magnetic Field Laboratory, CNRS and MPI-FKF,
BP 166, 38042 Grenoble Cedex 09, France}
\address {$^{2)}$ Institut Universitaire de France and Universit\'e J. Fourier, BP41,
F-38400 St. Martin d'H\`eres, France}
\address {$^{3)}$ Laboratoire de Spectrom\'etrie Physique, Universit\'e
J. Fourier, BP87, F-38402 St. Martin d'H\`eres, France}
\address {$^{4)}$ Dipartimento di Chimica, Universit\`a di Perugia,
I-06100 Perugia, Italy}
\date {\today}
\maketitle

\begin {abstract}

We present an extensive NMR study of the spin-1/2 antiferromagnetic
Heisenberg ladder $\rm Cu_2(C_{5}H_{12}N_{2})_2Cl_4$
in a magnetic field range 4.5 - 16.7T.
By measuring the proton NMR relaxation rate $1/T_{1}$ and
varying the magnetic field around the critical field
$H_{c1}=\Delta / g \mu_{B} \approx 7.5$T,
we have studied the transition from a gapped spin liquid ground state
to a gapless magnetic regime which can be described as a Luttinger liquid.
We identify an intermediate regime $ T \geq | H - H_{c1} |$, where the
spin dynamics is (possibly) only controlled by the T=0
critical point $H_{c1}$.

\end {abstract}


]
\narrowtext

One of the most fascinating effect in quantum magnetism is perhaps the
possibility to realize a T=0 phase transition \cite{Hertz76}
from a "quantum disordered" (gapped spin liquid) ground state to a
Luttinger liquid state in one dimension (1D) \cite{Haldane80} or to a 
N\'eel ordered state in two dimensions (2D) \cite{2DHAF,Chubukov93}.
There are well known examples of
gapped spin liquids in 1D: for integer-spin Heisenberg antiferromagnetic 
(HAF) chains \cite{Haldane83,Schulz86} or spin-1/2 HAF even-leg ladders
\cite{Dagotto96}, quantum fluctuations induce a spin gap $\Delta$
between a singlet (S=0) ground state and triplet (S=1) excited states.
An external magnetic field lifts the triplet degeneracy and induces a
second order T=0 phase transition at a critical field $H_{c1} \equiv
\Delta$ when the lowest branch of the triplet crosses the ground state.
At this critical field, dynamical properties are defined by universal
exponents \cite{Affleck90,Sachdev94,Chitra97}.
Above $H_{c1}$, the ground state is magnetic with an
algebraic decay of the correlation functions.
Close to $H_{c1}$, the low frequency ($\omega < T$) spin dynamics should be
governed in an universal way by the T=0 critical point $H_{c1}$. 
In the vicinity of this point, divergent quantum correlations are
cut-off by thermal fluctuations at a length $\xi_T$ which become the 
only relevant length scale.  This is the {\it quantum critical} regime 
(QCR) where the temperature is the only energy scale
over the large intermediate region  $J > T > |H-H_{c1}|$
\cite{Chubukov93}.  The true
challenge to experiments is to
observe in a single system, all the sequences of the $T \rightarrow 0$
regimes as well as the finite temperature critical regime.

In a gapped HAF, such quantum phase transition can be experimentally
studied only if the gap $\Delta$ is comparable to accessible magnetic
fields.  The situation is hopless in the S=1/2 spin ladder compound $\rm
SrCu_{2}O_{3}$ ($\Delta \geq 400$K) \cite{SrCuOpaper}.  On the other
hand, it is now well established that the organo-metallic compound $\rm
Cu_2(C_{5}H_{12}N_{2})_2Cl_4$ \cite{Chiari90} is a unique representative
of HAF S=1/2 ladders with small exchange constants ($J_{\perp}\approx
13.2$K, $J_{\parallel} \approx 2.5$K) and a spin gap ($\Delta_{k=\pi}
\simeq J_{\perp} - J_{\parallel} = 10.5 \pm 0.3$K)
\cite{Hammar96,Chabous97,Hayward96,Weihong97} which makes the entire
phase diagram experimentally accessible.

This Letter describes a proton ($\rm ^{1}H$) NMR study of $\rm
Cu_2(C_{5}H_{12}N_{2})_2Cl_4$ in magnetic fields ranging from 4.5 to
16.7 Tesla.  The complete phase diagram (including the various
temperature crossovers) has been observed experimentally.  We
unambiguously identify three different regimes when $T \rightarrow 0$:
1) a {\it gapped phase} for $H < H_{c1} \approx 7.5$T, defined by a spin
liquid ground state, where the energy gap, deduced from nuclear
spin-lattice relaxation time ($T_{1}$) measurements, is linearly reduced
by the magnetic field; 2) a {\it magnetic phase} with a gapless ground
state for $H_{c1} < H < H_{c2} \approx 13.2$T, characterized by a
power-law divergence of $1/T_{1}$ consistent with an interpretation
based on fermions in one dimension.  3) a {\it fully polarized gapped
phase} above $H_{c2}$.  At intermediate temperatures, we analyze the
crossovers between these regimes and give for the first time 
convincing evidences that a quantum critical regime is observed when
$\Delta > T > | H - H_{c1}|$.

Proton NMR experiments have been performed with the magnetic field
direction along the $\vec{b}$ axis (perpendicular to the [101] ladder
direction) of small single crystals (typically 100-200$\mu g$ each).
We used conventionnal pulse spin-echo sequences and frequency-shifted
(summed Fourier transform) processing \cite{Clark95}.
In a first run ($H < 8.7$T), a set of five crystals with their
$\vec{b}$ axis oriented has been used; for larger fields,
{\it one} single crystal was utilised.
While spectra displayed minor variations, the absolute
values of $T_{1}$ were found reproducible between these two runs.
A typical proton NMR spectra is displayed in Fig. \ref{fig_shift}.
Note that the large magnetization at 16.7 Tesla allows to resolve all 24
proton sites.  However, since the total width of the spectra follows the
magnetization $M$, this discrimination becomes increasingly difficult as
$M \rightarrow 0$.  Arrow in Fig.  \ref{fig_shift} indicates the line
(I) where the hyperfine shift $K$ and $T_{1}$ have been measured (see
Ref.  \cite{chab_rmn} for a discussion of the proton sites assignment).

The temperature dependence of the magnetic hyperfine shift $K \sim
\langle S^{z}_{i}\rangle /H$ at various magnetic fields is shown in Fig.
\ref{fig_shift}.  The uniform static ($q=0,\omega=0$) susceptibility
$\chi_{0} = M/H$ at different fields is also displayed on the same scale
using the relation $K = (A_{I} / g_{b} \mu_{B}) \chi_{0}$ with the
hyperfine coupling $A_{I} \approx 2.9$kOe and the g-factor $g_{b} =
2.03$ \cite{Chabous97}.  For $\rm H < H_{c1}$, both $K$ and $\chi_{0}$
drop exponentially due to the effective gap $\Delta_{h} = \Delta -
g\mu_{B}H$ between the singlet and the lowest triplet state.

As the critical field $\rm H_{c1}$ is crossed, {\it both} $K$ and
$\chi_{0}$ show a clear persistent downturn at low temperatures due to
residual short range antiferromagnetic correlations in the intermediate
phase.  When $T \rightarrow 0$, $K$ and $\chi_{0}$ go to a finite
value, which rises continuously as the field is increased from $H_{c1}$ 
to $H_{c2}$, following the increase of ground state 
magnetization.  Above $H_{c2}$ where $M_{(T=0)}$ is saturated
\cite{Chabous97}, $M/H$ should decreases as $1/H$ ($H
>H_{c2}$, $T \rightarrow 0$).  This is already visible at
$T\approx1.4$K, where $K$ at 16.7T is smaller than at 15T.  The
maximum value of $K$, extrapolated to $T=0$ corresponds to
$\langle S^{z} \rangle = \frac{1}{2}$.  Above $T \approx J_{\perp}$, we 
recover a Curie-like tail for all fields, which defines the "classical"
(decorrelated) regime.

We now discuss the spin-lattice relaxation rate $1/T_{1}$
which is expressed in terms of the magnetic structure
factors \cite{Moriya56}
\begin{eqnarray}
\frac{1}{T_1} & = & \frac{(\gamma_{n}\gamma_{e}\hbar)^{2}}{2} \sum_{q,\alpha}
{ F_{\alpha} (q) \cdot S_{\alpha} (q,\omega_{n})} \nonumber \\
S_{\alpha} (q, \omega_{n}) & = & \int \exp(i\omega_{n}t) dt 
\langle S_{\alpha}(q,t)  S_{\alpha}(-q,0) \rangle \; , \label{equ_1}
\end{eqnarray}
where $F_{\alpha} (q)$ are hyperfine form factors of dipolar
origin \cite{Henkens78} and $\alpha = z, \pm $ represents the
longitudinal and transverse components, respectively.

At low fields and low temperatures, $T \ll H_{c1} - H$, ({\it singlet
gapped phase}), $1/T_{1}$ falls off exponentially with a characteristic
energy gap $\Delta_{h} = \Delta - g\mu_{B}H$ [see Fig.  2(a)].  This
effective gap is represented in the lower graph of Fig.  2(a).  This
result can easily be understood: for line (I), it has been shown
in a previous study \cite{chab_rmn} that two-magnon processes (near
$k=\pi$, $q \simeq 0$) in the "intrabranch" channel \cite{Sagi96}
dominate the nuclear relaxation.  It follows that $1/T_{1}$, in this
gapped phase, is driven by longitudinal correlations $S_{zz} (q \simeq
0,\omega_{n}) \sim \exp(-\frac{\Delta_{h}}{T})$.  As we approach the
critical field, $T \approx H_{c1} -H $, the analysis breaks down since
competing mechanisms like "staggered" direct, $q=\pi$, processes take
over and dominate the relaxation \cite{Sagi96}.  At higher
temperatures, $T \geq \Delta$, $1/T_{1}$ is constant, as expected in the
classical limit \cite{Moriya56}.

In the {\it magnetic phase} ($H_{c1} < H < H_{c2}$) \cite{Chabous97},
$1/T_{1}$ turns upward at low temperature in striking contrast
with its behavior in the gapped phase [see top parts of Fig. 2(a) and 2(b)].
Very close to $H_{c1}$, a divergence is readily visible below
2-3K.  As the magnetic field is increased just above $H_{c1}$
($7.5<H<9$T), the
divergence becomes more pronounced and develops at temperatures below 
5K. At higher fields, the divergent behavior is replaced by a smooth
increase as the temperature is lowered below 10K and progressively
vanishes as we approach the upper critical field $H_{c2}$.

To conclude the $T=0$ limit, we discuss the "high field phase" which
appears above $H_{c2} = J_{\perp} + 2J_{\parallel}$ \cite{Chabous97}.
It is apparent [bottom of Fig.  2(b)] that the relaxation rate {\it decays
exponentially} at low temperatures with an activation energy
$\Delta^{up} = g \mu_{B} ( H - H_{c2})$.  Since the T=0 magnetization is
saturated above $H_{c2}$, the spin system is fully polarized
\cite{Chabous97} and all dimers are in the triplet state
$|$$\uparrow\uparrow \rangle$.
$\Delta^{up}$ can therefore be interpreted as the energy gap between the 
"fully polarized"
ground state $| FGS \rangle = \Pi_{i}|$$\uparrow\uparrow \rangle_{i}$
and the lowest excited states generated by a single spin flip.
The balance between the energy gain $J_{\perp}+2J_{\parallel}$
when the antiferromagnetic couplings are satisfied and the Zeeman
term $g\mu_{B}H$ determines the value $g\mu_{B}(H - H_{c2})$
for the energy gap.

At this stage, we propose the experimental phase diagram schematically
shown in Fig.  \ref{resume}.  The $T=0$ phases discussed above are:  1)
the gapped spin liquid phase; 2) a magnetic phase to be characterized
(see below) and 3) the gapped polarized phase.

At finite temperatures, these phases are separated by crossover lines:
(A) and (B) in Fig. \ref{resume} correspond, respectively, to the onset
of the gapped spin-liquid regime and of the gapped polarized regime,
both characterized by an exponential decay of $1/T_{1}$.  Line (C) sets
the upper boundary of the "magnetic phase", where a divergent behavior
of $1/T_{1}$ [$\sim T^{-\alpha}$] is observed.  At higher temperatures
($T \geq \Delta$), we have the classical regime defined by $1/T_{1}
\approx cst$ \cite{Moriya56} and a Curie-like behavior of $K \sim 1/T$
at all magnetic fields [see Fig \ref{fig_shift}].  What can be said
about the large intermediate region between these crossover lines?
On one hand, $1/T_{1}$ is found to be nearly T-independent in the range
$6.6 < H < 9$T, above a characteristic temperature "$T_{0}$" $\sim | H -
H_{c1} |$ [see Fig. 2].  On the other hand, $K$ shows a clear field
dependence below $T < \Delta$.  This is no longer a
classical regime: we propose that this region corresponds to a QCR
where $1/T_{1}$ is predicted to be almost temperature independent
\cite{Chubukov93} while, at the same time, static properties
depart from the classical picture.

In the final part of this Letter, we discuss the nature of the
magnetic phase, particularly the origin of the low temperature
divergence of $1/T_{1}$ between $H_{c1}$ and $H_{c2}$.

Since a 3D field-induced ordering above $H_{c1}$ has been observed in
specific heat measurements below $T_{N} (H) \leq 0.8$K \cite{Hammar97},
we first consider critical fluctuations as a possible origin for the
observed divergence.  In this scenario, one expects a
divergence of $1/T_{1}$ {\it in a range} $\delta T \approx T_{N}$ above
$T_{N}$ \cite{Moriya62} with a generic behavior $T^{-1}_{1} \sim (T -
T_{N})^{-\nu}$ ($0.5 < \nu < 1$).  {\it The exponent $\nu$ must be
independent of the magnetic field}.  We cannot fit our data taking into
account the known field dependence of $T_{N}(H)$ \cite{Hammar97} without
releasing this constraint.  Moreover, $1/T_{1}$ starts to diverge at $T
\approx 5$K, which is at least $\delta T \approx 5-6T_{N}$ above
$T_{N}$.  From these arguments, it is clear that the onset of a 3D
ordering cannot explain alone the behavior of $1/T_{1}$ in this
temperature range ($T \geq 1.3$K).
Hence, 1D quantum fluctuations have to be invoked to explain
our results \cite{comment2}.

We then propose an analysis in terms of fermions in one dimension ({\it
i.e} a Luttinger liquid \cite{Haldane80}) in the regime $T \ll H -
H_{c1}$.  The spin-ladder Hamiltonian can be converted into a 1D
interacting spinless fermions model using a Jordan-Wigner transformation
\cite{Haldane81,Schulz86,Chitra97}.  The spectrum consists of two
bands with energies $\hbar \omega_{k} = \pm ( \Delta +
\frac{c^{2}}{2\Delta}k^{2} )$.  In this mapping, the magnetic field $H$
plays the role of the chemical potential $\mu = g \mu_{B} (H - H_{c1})$ 
of the fermions (at $H=0$, $\mu$ lies in the middle of the spectrum).
Above $H_{c1}$ ($\mu >0$), the density of fermions $n$ increases
with $\mu$ as $n \propto M \sim \sqrt{\mu}$ \cite{Schulz80,Sachdev94}.
Since there is no gap, direct nuclear relaxation processes are allowed
and one expects an enhancement of the relaxation by an amount related to the
magnetization ($M \propto n$) of the electronic system.  Within this
picture, the staggered part $S_{\perp} (q\approx\pi, \omega_{n})$ leads
to $1/T_{1} \sim T^{-\alpha}$ ($\alpha=0.5$) when $H \rightarrow
H^{+}_{c1}$ \cite{Schulz86,Sagi96}.  At higher fields, $n$ increases and
interactions give rise to non-universal behaviors of the spin
correlations, {\it i.e} dependent of the microscopic details.  For
instance, Ref.  \cite{Chitra97} predicts that $\alpha$ stays close to
$0.5$ above $H_{c1}$ for a ladder in contrast with other S=1/2 gapped
systems.

This picture agrees with our results, at least below 9 Tesla, and
becomes poorer above due, possibly, to large interactions between
excitations \cite{Chabous97,chab_rmn}.  The data in Fig. 2 show that
the divergent term (controlled by the exponent $\alpha$), increases from
$H_{c1}$ and is maximum around $H \approx8.5-9$T.  When $H \rightarrow
H_{c2}$, the divergence weakens and the exponent $\alpha$ cannot be
reliably estimated.  An important point is that the exponent $\alpha$ is
related to the exponent $\eta(=1-\alpha)$ controlling the decay of the
spatial correlation ,$\langle S_{0}S_{r} \rangle \sim (-1)^{r}
|r|^{-\eta}$.  Our data are roughly consistent with exact
diagonalization calculations of $\langle S_{0}S_{r} \rangle$ for Haldane
chains in the gapless phase \cite{Sakai91}:  both $\eta$ and $\alpha$
are close to 0.5 at $H_{c1}$ and $H_{c2}$.  In between, $\eta$ should
have a minimum value $\eta \approx 0.3$ \cite{Sakai91}, meaning a
maximum of $\alpha$.  To summarize, we have shown that the field
dependent $1/T_{1}$ divergence is a 1D effect and not an onset of the
3D ordering which occurs at lower temperature.  Even though the exponent
$\alpha$ cannot be estimated precisely, a value of $\alpha=0.5$ at
$H_{c1}$ is consistent with our data.

In conclusion, the magnetic field-temperature phase diagram
of the spin-1/2 HAF ladder $\rm Cu_2(C_{5}H_{12}N_{2})_2Cl_4$
has been {\it completely} explored by probing the low frequency
spin dynamics.
Three $T \rightarrow 0$ regimes are identified.
1) a gapped spin liquid phase for $H < H_{c1}$;
2) a Luttinger liquid phase, well defined for $H_{c1} \leq H$,
characterized by a field-dependent correlation exponent $\alpha$.
3) a fully polarized gapped phase $H > H_{c2}$.
We emphasize that the exponent $\alpha$ derived from our data
is in qualitative agreement with a Luttinger liquid picture very
close to $H^{+}_{c1}$. Nevertheless, further studies of the field 
dependence of the exponent $\alpha$ are needed.  Finally, various
temperature crossovers are observed and closely resembles theoretical
predictions for a "quantum critical" regime
in a range $\Delta > T > | H - H_{c1} |$.

We thank R. Chitra, T. Giamarchi, D.K. Morr, D. Poilblanc and S. Sachdev for
stimulating discussions. One of us (M.E.H.) has received financial support
from a Bourse Chateaubriand of the Minist\`ere fran\c{c}ais
des Affaires \'etrang\`eres.

\begin{figure}
\caption{Top panel :  $\rm ^{1}H$ NMR spectrum at $f_{0} = 710.1$MHz and
T=5.1K.  All measurements reported here were made on the line marked by
an arrow.  Bottom panel:  Magnetic hyperfine shift $K_{I}$ as a function
of H and T. Bold and dotted lines represent macroscopic susceptibility
data obtained by AC-SQUID magnetometer (h=0.1T) and torsional
oscillator magnetometer, respectively \protect\cite{Crowell96}.  The
normalization factor gives an hyperfine coupling $A_{I} \approx
2.9$kOe.}
\label{fig_shift}
\end{figure}

\begin{figure}                                                                           
\caption {Panel (a): Temperature dependence of $1/T_{1}$
through the critical field $H_{c1}$.
The lower part covers the singlet gapped phase ($H < H_{c1}$)
while the upper part is in the magnetic phase above $H_{c1}$.
The dahsed lines are guides to the eyes.
Lines in the inset are low-T fits to the field
dependent energy gap $\Delta_{h} = \Delta - g \mu_{B} H$ with
$g_{b}=2.03$. Experimental values of $\Delta_{h}$ are shown
in the lower graph ($\Box$) together with those obtained
from $\chi (T)$ and $M(H)$ ($\bullet$) \protect\cite{Chabous97}.
Lines labeled (A) and (C) correspond to the crossover lines of Fig. \protect\ref{resume}.
Panel (b): Temperature dependence of $1/T_{1}$ through the critical field $H_{c2}$.
The upper part is in the magnetic phase as $H_{c2}$ is approached
while the lower part is in the "fully polarized" gapped phase
($H > H_{c2}$).
Lines in the inset are low-T fits with an energy gap
$\Delta^{up} = g \mu_{B} (H - H_{c2})$ and $g_{b}=2.03$.
Experimental values of $\Delta^{up}$ are shown in the
lower graph ($\Box$) together with $H_{c2}$ ($\bullet$)
\protect\cite{Chabous97}.}
\end{figure}

\begin{figure}
\caption{H-T phase diagram which can be tentatively drawn from the
present experiment.  Dashed lines represent crossovers between the
different regimes discussed in the text:  line (D) is the QCR-classical
crossover, lines (A) and (B) correspond to $T \approx H_{c1} -H$ and
$T \approx H_{c2} -H$, respectively.  Line (C) is the QCR to Luttinger
liquid (LL) regime crossover.  The black region corresponds to the 3D
ordered phase with $T_{N}(max) \approx 0.8$K \protect \cite{Hammar97}.}
\label{resume}
\end{figure}

\end {document}
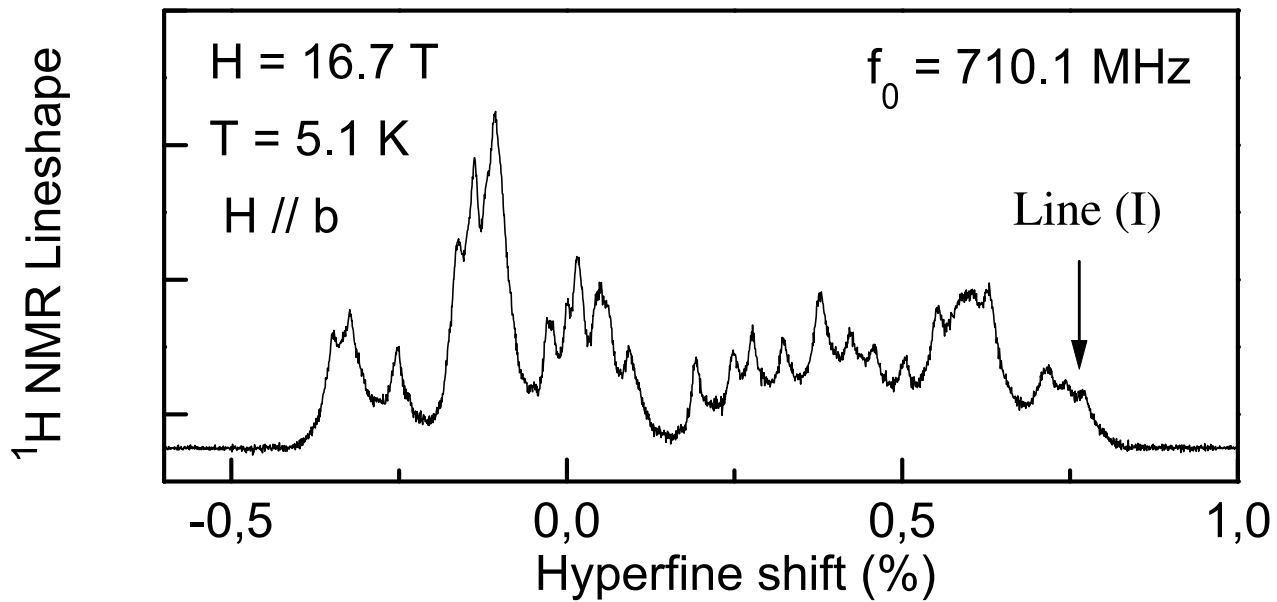
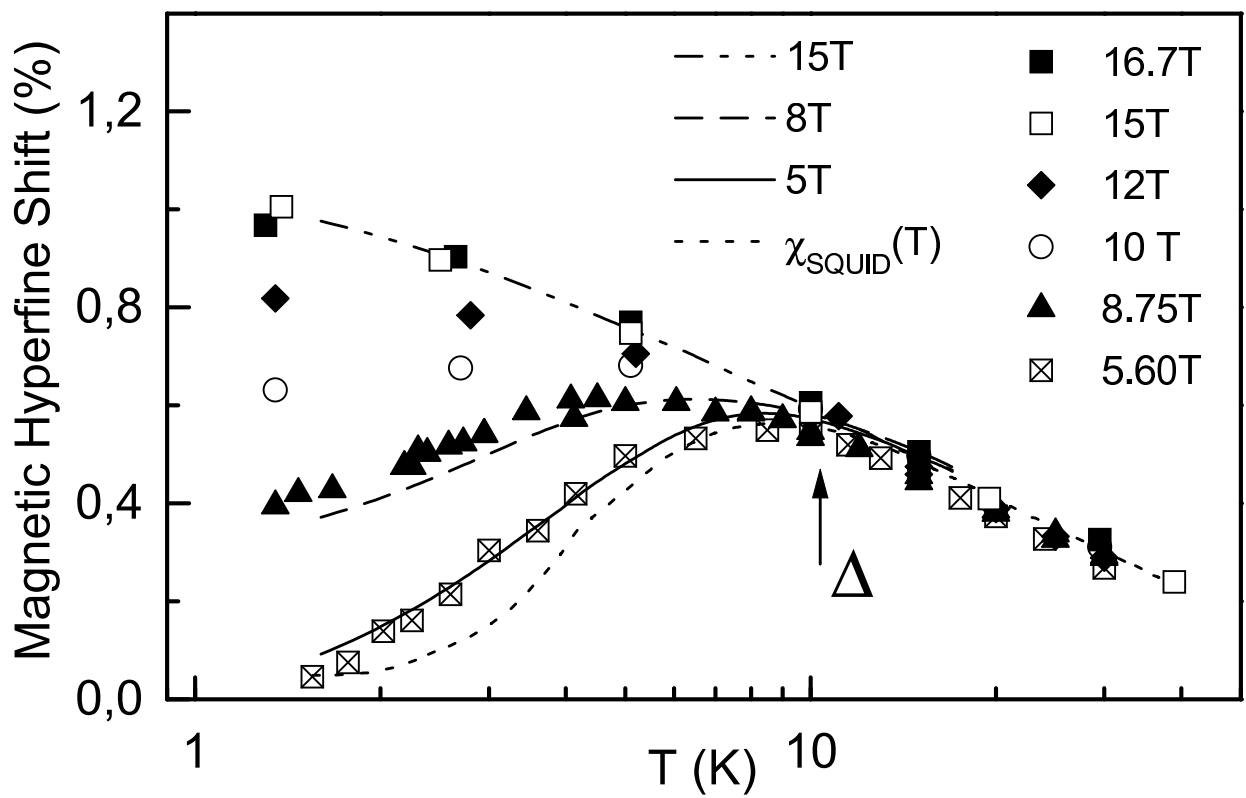
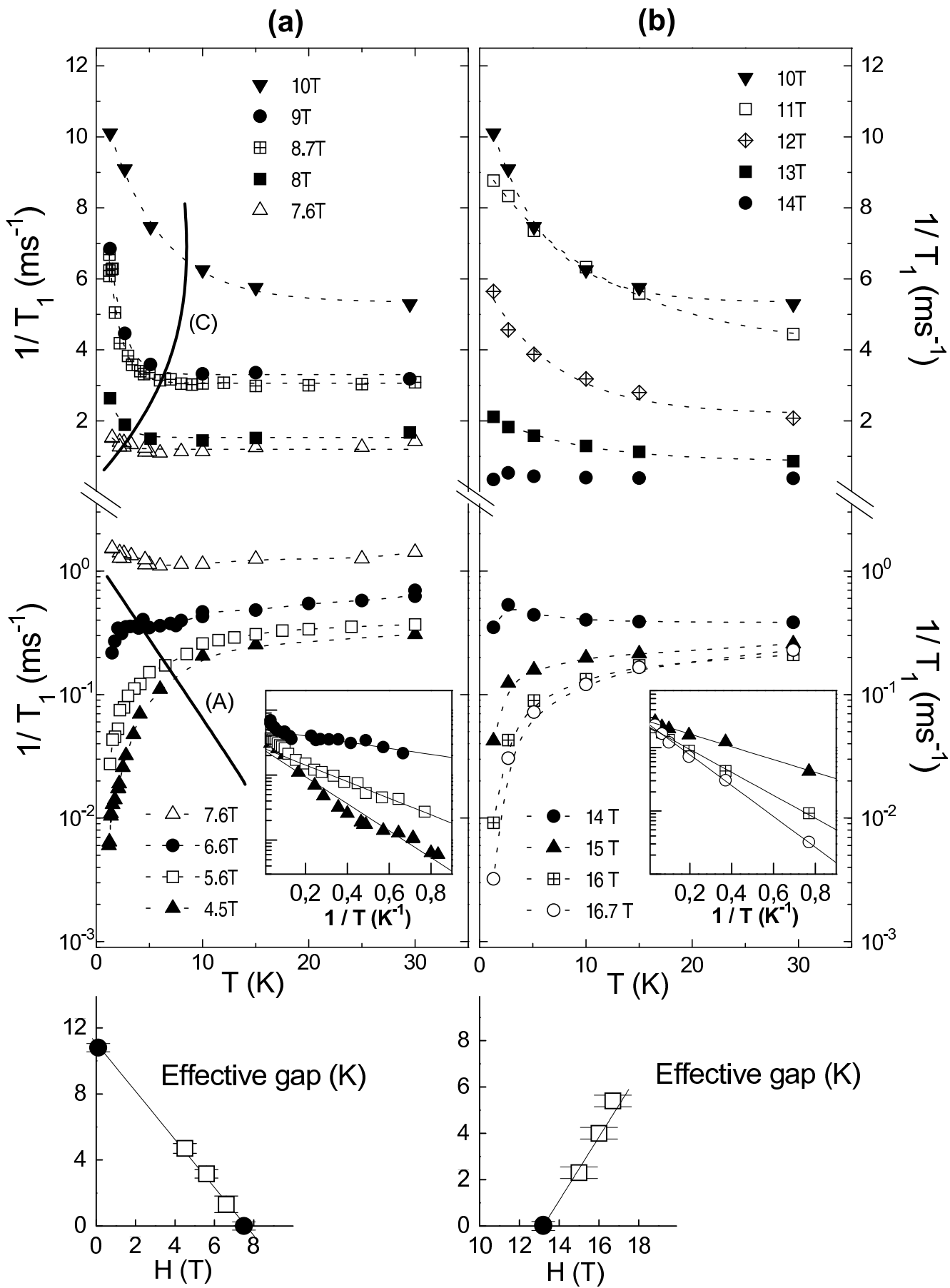
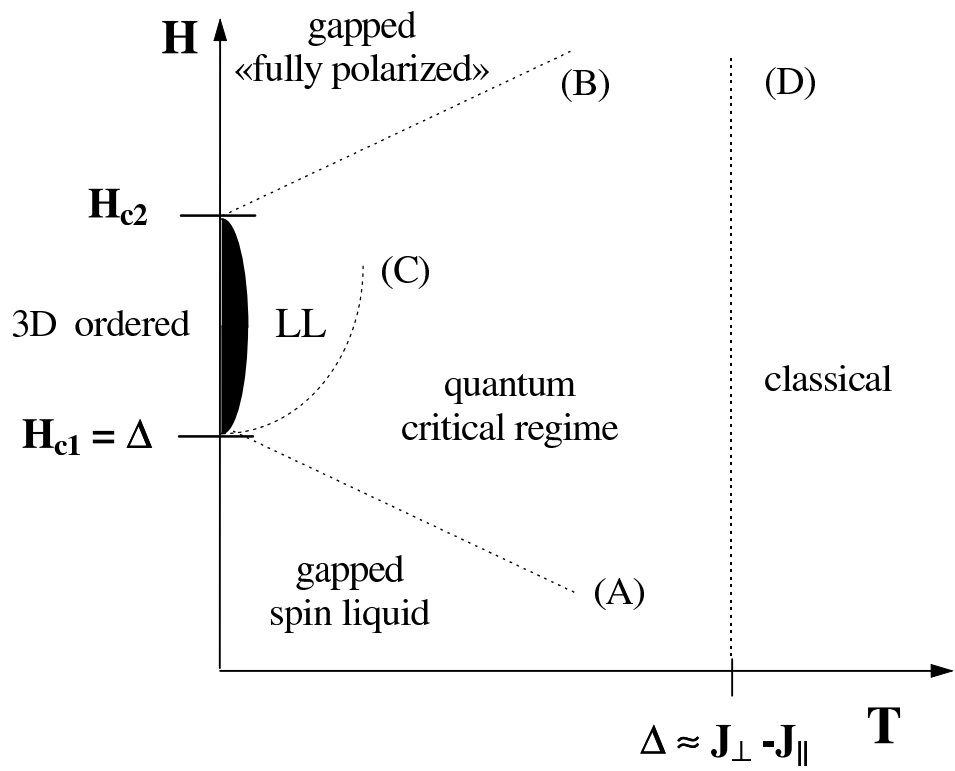
